\documentclass[letterpaper, 10 pt, conference]{ieeeconf}  

\IEEEoverridecommandlockouts                              

\usepackage{amssymb}
\usepackage{amsmath}
\usepackage{bm}
\usepackage{mathrsfs}
\usepackage{comment} 
\usepackage{algorithmic,algorithm}
\usepackage{shuffle}
\usepackage{graphicx} 
\usepackage{caption}
\usepackage{lipsum}
\usepackage{color}
\usepackage{enumerate}
\usepackage{amsfonts}
\usepackage{subfigure} 
\usepackage{array}
\usepackage{enumerate}
\usepackage{setspace}
\usepackage{multirow}

\title{\LARGE \bf
Identification of System Vulnerability under a Smart Sensor Attack via Attack Model Reduction*
}

\author{Ruochen Tai$^{1}$, Liyong Lin$^{1}$ and Rong Su$^{1}$
\thanks{*The research of the project was supported by Ministry of Education, Singapore, under grant AcRF TIER 1-2018-T1-001-245 (RG 91/18).}
\thanks{$^{1}$The authors are affliated with Nanyang Technological University, Singapore. (Email: {\tt\small ruochen001@e.ntu.edu.sg; llin5@e.ntu.edu.sg; rsu@ntu.edu.sg}).
        }%
}

\begin{document}

\maketitle
\thispagestyle{empty}
\pagestyle{empty}

\begin{abstract}

In this work, we investigate how to make use of model reduction techniques to identify the vulnerability of a closed-loop system, consisting of a plant and a supervisor, that might invite attacks. Here, the system vulnerability refers to the existence of key observation sequences that could be exploited by a specific smart sensor attack to cause damage infliction. We consider a nondeterministic smart attack, i.e., there might exist more than one attack choice over each received observation, and adopt our previously proposed  modeling framework, where such an attack is captured by a standard finite-state automaton. For a given supervisor $S$ and a smart sensor attack model $A$, another smart attack model $A'$ is called attack equivalent to $A$ with respect to $S$, if the resulting compromised supervisor, defined as the composition of the supervisor $S$ and attack model $A'$, is control equivalent to the original compromised supervisor, defined as the composition of $S$ and $A$. Following the spirit of supervisor reduction that relies on the concept of control congruence, we will show that, this problem of synthesizing a reduced smart attack model $A'$ that is attack equivalent to $A$ with respect to $S$, can be transformed to a classical supervisor reduction problem, making all existing synthesis tools available for supervisor reduction directly applicable to our problem. A simplified and ideally minimum-state attack model can reveal all necessary observation sequences for the attacker to be successful, thus, reminds system designers to take necessary precautions in advance, which may improve system resilience significantly. An example is presented to show the effectiveness of our proposed attack model reduction technique to identify the system vulnerability.

\end{abstract}


\section{INTRODUCTION}
\label{sec:intro}

As an integration of cyber information and physical world, cyber-physical systems (CPS) has been playing a much more significant role in the modern society due to its precise control, remote collaboration and autonomous functions. The realization of these powerful features heavily relies on the network system (cyber part), which might be compromised and taken use of to cause irreparable damage by malicious attackers. 
Recently, the security issue of cyber-physical systems has drawn a lot of attention from both the computer science community and the systems control community. Quite a few works have been devoted to the cyber security issues related to control, optimization, and computation \cite{CAS2008}-\cite{ZCSC2012}. 

In the DES community, plenty of studies on security issues have emerged and could be categorized into three classes: 1) attack detection and security verification \cite{CarvalhoEnablementAttacks}-\cite{WP}, 2) synthesis of attackers \cite{Goes2017}-\cite{ZSL2021}, and 3) synthesis of resilient supervisors \cite{Su2018}, \cite{Su20}-\cite{LS20BJ}. For the synthesis of attackers, the previous works target to find a set of attack sequences that could provide the attackers with a specific attack strategy of implementing attacks on sensors and actuators such that damage infliction could be caused on the system. However, there usually exist multiple attackers that could successfully lead the system to a damage state and the methods adopted in the previous works might result in the synthesis of attackers with numerous states and transitions, making it quite difficult for the system designers to intuitively discover the system flaws in terms of security and then designing efficient defending strategies. Thus, motivated by the advantage of a simplified and intuitive attack strategy, in this work, we study how to identify the system vulnerability under a smart sensor attack by designing an attack model reduction technique. Here, the system vulnerability refers to those key observation sequences fired by the plant, which could be taken use of by the attacker to determine attack actions. The contribution of this work is: We design an efficient attack model reduction method that could potentially compute the minimum-state attack model, which could reveal all necessary observation sequences for the attacker to be successful. Such a technique could assist system designers in quickly understanding and pinpointing the weaknesses that might bring about security risks, and designing precautionary measures, e.g., synthesizing resilient supervisors, to prevent the malicious attackers from disturbing the system operations, and thus improve the resilience.



This paper is organized as follows. In Section \ref{sec:preliminaries}, we recall the preliminaries which are needed for understanding this paper. In Section \ref{sec:model reduction}, we introduce the system setup and present the model constructions, based on which the attack model reduction technique is presented. An example to show the effectiveness of the proposed method is given in Section \ref{sec:example}. Finally, the conclusions are drawn in Section \ref{sec:conclusion}.


\section{PRELIMINARIES}
\label{sec:preliminaries}

Given a finite alphabet $\Sigma$, let $\Sigma^{*}$ be the free monoid over $\Sigma$ with the empty string $\varepsilon$ being the unit element and the string concatenation being the monoid operation. For a string $s$, $|s|$ is defined as the length of $s$. Given two strings $s, t \in \Sigma^{*}$, we say $s$ is a prefix substring of $t$, written as $s \leq t$, if there exists $u \in \Sigma^{*}$ such that $su = t$, where $su$ denotes the concatenation of $s$ and $u$. A language $L \subseteq \Sigma^{*}$ is a set of strings. The prefix closure of $L$ is defined as $\overline{L} = \{u \in \Sigma^{*} \mid (\exists v \in L) \, u\leq v\}$. 
The event set $\Sigma$ is partitioned into $\Sigma = \Sigma_{c} \dot{\cup} \Sigma_{uc} = \Sigma_{o} \dot{\cup} \Sigma_{uo}$, where $\Sigma_{c}$ (respectively, $\Sigma_{o}$) and $\Sigma_{uc}$ (respectively, $\Sigma_{uo}$) are defined as the sets of controllable (respectively, observable) and uncontrollable (respectively, unobservable) events, respectively.  As usual, $P_{o}: \Sigma^{*} \rightarrow \Sigma_{o}^{*}$ is the natural projection defined such that
\begin{enumerate}[(1)]
\setlength{\itemsep}{3pt}
\setlength{\parsep}{0pt}
\setlength{\parskip}{0pt}
    \item $P_{o}(\varepsilon) = \varepsilon$,
    \item $(\forall \sigma \in \Sigma) \, P_{o}(\sigma)=
    \left\{
    \begin{array}{rcl}
    \sigma       &      & {\sigma \in \Sigma_{o},}\\
    \varepsilon  &      & {\rm otherwise,}
    \end{array} \right.$
    \item $(\forall s \in \Sigma^*, \sigma \in \Sigma) \, P_{o}(s\sigma) = P_{o}(s)P_{o}(\sigma)$.
\end{enumerate}
A finite state automaton $G$ over $\Sigma$ is given by a 5-tuple $(Q, \Sigma, \xi, q_{0}, Q_{m})$, where $Q$ is the state set, $\xi: Q \times \Sigma \rightarrow Q$ is the (partial) transition function, $q_{0} \in Q$ is the initial state, and $Q_{m}$ is the set of marker states. 
We write $\xi(q, \sigma)!$ to mean that $\xi(q, \sigma)$ is defined and also view $\xi \subseteq Q \times \Sigma \times Q$ as a relation. $En_{G}(q) = \{\sigma \in \Sigma|\xi(q, \sigma)!\}$.
$\xi$ is also extended to the (partial) transition function $\xi: Q \times \Sigma^{*} \rightarrow Q$ and the transition function $\xi: 2^{Q} \times \Sigma \rightarrow 2^{Q}$ \cite{wonham2015supervisory}, where the later is defined as follows: for any $Q' \subseteq Q$ and any $\sigma \in \Sigma$, $\xi(Q', \sigma) = \{q' \in Q|(\exists q \in Q')q' = \xi(q, \sigma)\}$. 
Let $L(G)$ and $L_{m}(G)$ denote the closed-behavior and the marked behavior of $G$~\cite{wonham2015supervisory}, respectively. When $Q_{m} = Q$, we shall also write $G = (Q, \Sigma, \xi, q_{0})$ for simplicity. We denote by $|Q|$ the size of the state set $Q$. When the state set is not explicitly mentioned, we also write $|G|$ for the size of an automaton, namely the size of its state set.

As usual, for any two finite state automata $G_{1} = (Q_{1}, \Sigma_{1}, \xi_{1}, q_{1,0}, Q_{1,m})$ and $G_{2} = (Q_{2}, \Sigma_{2}, \xi_{2}, q_{2,0}, Q_{2,m})$, where $En_{G_{1}}(q) = \{\sigma|\xi_{1}(q, \sigma)!\}$ and $En_{G_{2}}(q) = \{\sigma|\xi_{2}(q, \sigma)!\}$, their synchronous product \cite{CassandrasDES2008} is denoted as $G_{1}||G_{2} := (Q_{1} \times Q_{2}, \Sigma_{1} \cup \Sigma_{2}, \zeta, (q_{1,0}, q_{2,0}), Q_{1,m} \times Q_{2,m})$, where the (partial) transition function $\zeta$ is defined as follows: for any $(q_{1}, q_{2}) \in Q_{1} \times Q_{2}$ and $\sigma \in \Sigma$:
\[
\begin{aligned}
& \zeta((q_{1}, q_{2}), \sigma) := \\ & \left\{
\begin{array}{lcl}
(\xi_{1}(q_{1}, \sigma), \xi_{2}(q_{2}, \sigma))  &      & {\rm if} \, {\sigma \in En_{G_{1}}(q_{1}) \cap En_{G_{2}}(q_{2}),}\\
(\xi_{1}(q_{1}, \sigma), q_{2})       &      & {\rm if} \, {\sigma \in En_{G_{1}}(q_{1}) \backslash \Sigma_{2},}\\
(q_{1}, \xi_{2}(q_{2}, \sigma))       &      & {\rm if} \, {\sigma \in En_{G_{2}}(q_{2}) \backslash \Sigma_{1},}\\
{\rm not \, defined}  &      & {\rm otherwise.}
\end{array} \right.
\end{aligned}
\]

For a plant $G$ modelled as a deterministic finite state automaton $G = (Q, \Sigma, \xi, q_{0}, Q_{m})$, a (feasible) supervisor of $G$ under $P_{o}: \Sigma^{*} \rightarrow \Sigma_{o}^{*}$ is a finite state automaton $S$ such that the controllability and observability constraints \cite{wonham2015supervisory} are satisfied on the closed-loop system behaviors $L(G||S)$. 

\emph{Supervisor reduction}: For a plant $G$, there may exist more than one supervisor that achieves a control objective, e.g., ensures that the closed-loop system behavior is contained in a predefined requirement language $E \subseteq \Sigma^{*}$. Two supervisors $S_{1}$ and $S_{2}$ of $G$ are control equivalent \cite{SW2004reduction} if $L(G||S_{1}) = L(G||S_{2})$ and $L_{m}(G||S_{1}) = L_{m}(G||S_{1})$. Let $\mathcal{F}(G,S)$ be the collection of all feasible supervisors of $G$ under partial observation $P_{o}$, which are control equivalent to a given supervisor $S$. It is desirable to find one supervisor $S_{*} \in \mathcal{F}(G,S)$ such that for all $S' \in \mathcal{F}(G,S)$ we have $|S_{*}| \leq |S'|$, i.e., the supervisor $S_{*}$ has the minimum number of states. It has been shown in \cite{SW2004reduction} that, unfortunately, finding $S_{*}$ based on the concept of control covers is NP-hard, even for a supervisor under full observation. Each control cover is a collection of subsets of states in $S$, in which the states of each subset are “control consistent”. Thus, by grouping those compatible states of $S$ together, we may get a new reduced supervisor $S'$ such that (1) $S'$ is control equivalent to $S$; (2) $|S'| < |S|$. For more details on supervisor reduction, We refer readers to \cite{SW2004reduction}-\cite{SW2018reductionJ}. 


\section{IDENTIFICATION OF SYSTEM VULNERABILITY VIA ATTACK MODEL REDUCTION}
\label{sec:model reduction}

In this section, we shall explain how to make use of model reduction techniques to identify the vulnerability of a closed-loop system, consisting of a plant and a supervisor, that might invite attacks. Our idea is: Firstly, we show the method of transforming a sensor attacker to a new supervisor for a new surrogate plant in the standard Ramadge-Wonham supervisory control problem \cite{LZS19}-\cite{LS20J}. Then, based on the transformed result, the attack model reduction problem is naturally reduced to the well-studied supervisor reduction problem.


\subsection{Component modelling under sensor attack}
\label{subsec:component modelling}

In supervisory control of discrete-event systems \cite{wonham2015supervisory}, the plant $G$ is under the control of a supervisor $S$ over some control constraint $(\Sigma_{c}, \Sigma_{o})$. However, considering the security issues in a cyber-threat environment, there might exist an attacker that could partially observe the system behaviors and carry out attacks to cause damage infliction on the plant. Specifically, in this work, we assume a sensor attacker is deployed in the closed-loop system to alter sensor readings such that the supervisor is deceived into issuing inappropriate control commands under fake sensor information. Next, we shall introduce how to model some components in supervisory control of discrete-event systems under sensor attack, mostly following the framework in \cite{LS20J}.

\textbf{Plant:} As usual, the plant $G$ is modelled as a finite state automaton $G = (Q, \Sigma, \xi, q_{0}, Q_{d})$, where $Q_{d} \subseteq Q$ is the set of damage states. Any state of $Q_{d}$ is a goal state that the sensor attack targets to induce $G$ to reach.

\textbf{Supervisor:} Under the absence of attacks, the supervisor $S$ is modelled as a finite state automaton $S = (Q_{s}, \Sigma, \xi_{s}, q_{s}^{init})$, where all states are marked.
The control command issued by the supervisor $S$ at state $q \in Q_{s}$ is defined to be $\Gamma(q) = En_{S}(q) = \{\sigma \in \Sigma|\xi_{s}(q,\sigma)!\} \in \Gamma = \{\gamma \subseteq \Sigma|\Sigma_{uc} \subseteq \gamma\}$, where $\Gamma$ is the set of control commands. We assume the supervisor $S$ will immediately issue a control command to the plant whenever an event $\sigma \in \Sigma_{o}$ is received or when the system initiates.

Next, we shall present a transformation construction procedure \cite{LS20J}, based on which we are able to view a sensor attacker as a new supervisor for a new surrogate plant. 

\textbf{Sensor attack constraints:} In this work, the sensor attacker is assumed to 1) implement replacement attack, 2) observe the events in $\Sigma_{o,a} \subseteq \Sigma_{o}$, where $\Sigma_{o}$ is the set of observable events of the plant, and 3) attack the events in $\Sigma_{s,a} \subseteq \Sigma_{o,a}$, where $\Sigma_{s,a}$ is the set of compromised events of the plant. We shall construct a model, named as sensor attack constraints $AC$ \cite{LS20,LS20J}, to capture the attack capabilities of the sensor attacker, which is presented as follows:
\[
AC = (Q_{ac}, \Sigma_{ac}, \xi_{ac}, q_{ac}^{init})
\]
\begin{itemize}
\setlength{\itemsep}{3pt}
\setlength{\parsep}{0pt}
\setlength{\parskip}{0pt}
    \item $Q_{ac} = \{q_{ac}^{init}, q^{obs}\}$
    \item $\Sigma_{ac} = \Sigma \cup \Sigma_{s,a}^{\#}$
    \item $\xi_{ac}: Q_{ac} \times \Sigma_{ac} \rightarrow Q_{ac}$
\end{itemize}
The (partial) transition function $\xi_{ac}$ is defined as follows:
\begin{enumerate}[1.]
\setlength{\itemsep}{3pt}
\setlength{\parsep}{0pt}
\setlength{\parskip}{0pt}
    \item For any $\sigma \in \Sigma - \Sigma_{s,a}$, $\xi_{ac}(q_{ac}^{init}, \sigma) = q_{ac}^{init}$.  
    \item For any $\sigma \in \Sigma_{s,a}$, $\xi_{ac}(q_{ac}^{init}, \sigma) = q^{obs}$. 
    \item For any $\sigma \in \Sigma_{s,a}$, $\xi_{ac}(q^{obs}, \sigma^{\#}) = q_{ac}^{init}$. 
\end{enumerate}
In the event set, any event $\sigma^{\#} \in \Sigma_{s,a}^{\#}$ is a relabelled copy of $\sigma \in \Sigma_{s,a}$, denoting the compromised event $\sigma$ sent by the sensor attacker. Such an event relabelling also implies that the supervisor could only observe $\Sigma_{s,a}^{\#}$ instead of $\Sigma_{s,a}$, which allows us to capture the sensor attack effects. The (partial) transition function $\xi_{ac}$ says that, the observation of any event in $\Sigma_{s,a}$ would lead to a transition to the state $q^{obs}$, denoted by Case 2, and then the sensor attacker may perform replacement attack, denoted by Case 3.
In the following text, we shall refer to $\mathscr{C}_{ac} = (\Sigma_{s,a}^{\#}, \Sigma_{o,a} \cup \Sigma_{s,a}^{\#})$ as the attacker's control constraint, that is, the sensor attacker could only disable events in $\Sigma_{s,a}^{\#}$ and observe events in $\Sigma_{o,a} \cup \Sigma_{s,a}^{\#}$, and $(\Sigma_{o,a}, \Sigma_{s,a})$ as the attack constraint.

\textbf{Transformed supervisor under attack:} We perform the next two steps to generate this model.

\emph{Step 1: Supervisor bipartization}. We shall firstly carry out a bipartization transformation on supervisor $S$ to explicitly encode the control command sending phase. For any supervisor $S = (Q_{s}, \Sigma, \xi_{s}, q_{s}^{init})$, the procedure to construct a bipartite supervisor $BT(S)$ \cite{LZS19}-\cite{LS20J} is given as follows:
\[
BT(S) = (Q_{bs}, \Sigma_{bs}, \xi_{bs}, q_{bs}^{init})
\]
\begin{enumerate}[1.]
\setlength{\itemsep}{3pt}
\setlength{\parsep}{0pt}
\setlength{\parskip}{0pt}
    \item $Q_{bs} = Q_{s} \cup Q_{s}^{com}$, where $Q_{s}^{com}:= \{q^{com} \mid q \in Q_s\}$
    \item $\Sigma_{bs} = \Sigma \cup \Gamma$
    \item \begin{enumerate}[a.]
        \setlength{\itemsep}{3pt}
        \setlength{\parsep}{0pt}
        \setlength{\parskip}{0pt}
            \item $(\forall q^{com} \in Q_{s}^{com}) \, \xi_{bs}(q^{com}, \Gamma(q)) = q$.
            \item $(\forall q \in Q_{s})(\forall \sigma \in \Sigma_{uo}) \, \xi_{s}(q, \sigma)! \Rightarrow \xi_{bs}(q, \sigma) = \xi_{s}(q, \sigma) =q$.
            \item $(\forall q \in Q_{s})(\forall \sigma \in \Sigma_{o}) \, \xi_{s}(q, \sigma)! \Rightarrow \xi_{bs}(q, \sigma) = (\xi_{s}(q, \sigma))^{com}$. 
        \end{enumerate}
    \item $q_{bs}^{init} = (q_{s}^{init})^{com}$
\end{enumerate}

\emph{Step 2: Attacked bipartite supervisor}. For a transformed bipartite supervisor $BT(S)$, due to the effects of event relabellings for $\Sigma_{s,a}$, we need to relabel any event $\sigma \in \Sigma_{s,a}$ to $\sigma^{\#}$, in order to reflect the receiving of the attacked copy $\sigma^{\#}$ instead of $\sigma$ at the supervisor side. The generated new model is denoted as bipartite supervisor under attack $BT(S)^{A}$ \cite{LZS19}-\cite{LS20J}, whose construction procedure is given as follows: 
\[
BT(S)^{A} = (Q_{bs,a}, \Sigma_{bs,a}, \xi_{bs,a}, q_{bs,a}^{init})
\]
\begin{enumerate}[1.]
\setlength{\itemsep}{3pt}
\setlength{\parsep}{0pt}
\setlength{\parskip}{0pt}
    \item $Q_{bs,a} = Q_{bs} \cup \{q^{no,covert}\} = Q_{s} \cup Q_{s}^{com} \cup \{q^{no,covert}\}$
    \item $\Sigma_{bs,a} = \Sigma \cup \Sigma_{s,a}^{\#} \cup \Gamma$
    \item \begin{enumerate}[a.]
        \setlength{\itemsep}{3pt}
        \setlength{\parsep}{0pt}
        \setlength{\parskip}{0pt}
        \item $(\forall q, q' \in Q_{s})(\forall \sigma \in \Sigma_{s,a}) \, \xi_{bs}(q, \sigma) = q' \Rightarrow \xi_{bs,a}(q, \sigma^{\#}) = q' \wedge \xi_{bs,a}(q, \sigma) = q$. 
        \item $(\forall q, q' \in Q_{bs})(\forall \sigma \in (\Sigma - \Sigma_{s,a}) \cup \Gamma) \, \xi_{bs}(q, \sigma) = q' \Rightarrow \xi_{bs,a}(q, \sigma) = q'$. 
        \item $(\forall q \in Q_{s})(\forall \sigma \in \Sigma_{s,a})\neg\xi_{bs}(q, \sigma)! \Rightarrow \xi_{bs,a}(q,\sigma^{\#}) = q^{no,covert}$.
        \item $(\forall q \in Q_{s})(\forall \sigma \in \Sigma_{o} - \Sigma_{s,a})\neg\xi_{bs}(q, \sigma)! \Rightarrow \xi_{bs,a}(q,\sigma) = q^{no,covert}$.
    \end{enumerate}
    \item $q_{bs,a}^{init} = q_{bs}^{init}$
\end{enumerate}
In the (partial) transition function $\xi_{bs,a}$, at Step 3.a, 1) all the transitions labelled by events in $\Sigma_{s,a}$ are replaced with the copies in $\Sigma_{s,a}^{\#}$, denoted by $\xi_{bs,a}(q, \sigma^{\#}) = q{'}$, and 2) the transitions labelled by events in $\Sigma_{s,a}$ and originally defined in $BT(S)$ at state $q$ would become self-loops since these events can be fired and are unobservable to the supervisor, denoted by $\xi_{bs,a}(q, \sigma) = q$. At Step 3.b, all the other transitions, labelled by events in $(\Sigma - \Sigma_{s,a}) \cup \Gamma$, are retained. Step 3.c and Step 3.d encode the covertness-broken situations, that is, the attacker is discovered \cite{LS20J}.

\textbf{Command execution:} 
The command execution automaton $CE$ is constructed to describe the phase from using a control command to executing an event at the plant \cite{LZS19}-\cite{LS20J}, which is given as follows:
\[
CE = (Q_{ce}, \Sigma_{ce}, \xi_{ce}, q_{ce}^{init})
\]
\begin{itemize}
\setlength{\itemsep}{3pt}
\setlength{\parsep}{0pt}
\setlength{\parskip}{0pt}
    \item $Q_{ce} = \{q^{\gamma}|\gamma \in \Gamma\} \cup \{q_{ce}^{init}\}$
    \item $\Sigma_{ce} = \Gamma \cup \Sigma$
    \item $\xi_{ce}: Q_{ce} \times \Sigma_{ce} \rightarrow Q_{ce}$
\end{itemize}
The (partial) transition function $\xi_{ce}$ is defined as follows:
\begin{enumerate}[1.]
\setlength{\itemsep}{3pt}
\setlength{\parsep}{0pt}
\setlength{\parskip}{0pt}
    \item For any $\gamma \in \Gamma$, $\xi_{ce}(q_{ce}^{init}, \gamma) = q^{\gamma}$. 
    \item For any $\sigma \in \gamma \cap \Sigma_{uo}$, $\xi_{ce}(q^{\gamma}, \sigma) = q^{\gamma}$. 
    \item For any $\sigma \in \gamma \cap \Sigma_{o}$, $\xi_{ce}(q^{\gamma}, \sigma) = q_{ce}^{init}$. 
\end{enumerate}

With such a command execution automaton and the bipartite supervisor construction procedure, the control equivalence could also be formulated as follows: two supervisors $S_{1}$ and $S_{2}$ are control equivalent if $L(G||CE||BT(S_{1})) = L(G||CE||BT(S_{2}))$ and $L_{m}(G||CE||BT(S_{1})) = L_{m}(G||CE||BT(S_{2}))$.

\textbf{Sensor attacker:} The sensor attack model is a finite state automaton $A = (Q_{a}, \Sigma_{a} = \Sigma \cup \Sigma_{s,a}^{\#}, \xi_{a}, q_{a}^{init})$ \cite{LS20,LS20J}, where all states are marked, then we have the closed-loop system is $G||CE||BT(S)^{A}||A$. The sensor attack model $A$ should satisfy: 1) the controllability and observability \cite{wonham2015supervisory} w.r.t. the control constraint $\mathscr{C}_{ac}$ are satisfied on the closed-loop system behaviors $L(G||CE||BT(S)^{A}||A)$, and 2) $L(A) \subseteq L(AC)$, that is, the sensor attacker should always follow the attack mechanism modelled by $AC$. In this work, the sensor attack model $A$ is known and we assume it is covert \cite{LS20,LS20J}, that is, the supervisor would not discover the sensor information inconsistency by comparison with the closed-loop system behavior in the absence of attack, which could be generated from models of the plant and supervisor. Our goal in this work is to compute a reduced attack model with a simplified attack logic for $A$ to reveal all the necessary observation sequences for successfully causing damage infliction. 


\subsection{Attack model reduction}
\label{subsec:attack model reduction}

Based on the component models presented in Section \ref{subsec:component modelling}, the closed-loop system under sensor attack is 
\[
B = G||CE||BT(S)^{A}||A = (Q_{b}, \Sigma_{b}, \xi_{b}, q_{b}^{init}, Q_{b,m})
\]
We could then view $G||CE||BT(S)^{A}$ as a new plant and the sensor attacker $A$ as a new supervisor over the control constraint $\mathscr{C}_{ac}$, which completes the transformation procedure from a sensor attacker to a new supervisor for a new surrogate plant in the standard Ramadge-Wonham supervisory control problem. 

\emph{Definition III.1 (Compromised supervisor under sensor attack)} Given a plant $G$, a supervisor $S$ and a sensor attacker $A$, $BT(S)^{A}||A$ is the compromised supervisor under sensor attack $A$ for $G$.

\emph{Definition III.2 (Attack equivalence)} Given a plant $G$, a supervisor $S$ and an attack constraint $(\Sigma_{o,a}, \Sigma_{s,a})$, two sensor attackers $A$ and $A'$ over $(\Sigma_{o,a}, \Sigma_{s,a})$ are attack equivalent on $(G,S)$ if $L(G||CE||BT(S)^{A}||A) = L(G||CE||BT(S)^{A}||A')$ and $L_{m}(G||CE||BT(S)^{A}||A) = L_{m}(G||CE||BT(S)^{A}||A')$, that is, two compromised supervisors under sensor attack $BT(S)^{A}||A$ and $BT(S)^{A}||A'$ are control equivalent on the plant $G$.




In this work, given a plant $G = (Q, \Sigma, \xi, q_{0}, Q_{d})$, a supervisor $S = (Q_{s}, \Sigma, \xi_{s}, q_{s}^{init})$, a sensor attacker $A = (Q_{a}, \Sigma_{a} = \Sigma \cup \Sigma_{s,a}^{\#}, \xi_{a}, q_{a}^{init})$ and an attack constraint $(\Sigma_{o,a}, \Sigma_{s,a})$, we need to find a reduced sensor attack model $A'$ such that $A$ and $A'$ over $(\Sigma_{o,a}, \Sigma_{s,a})$ are attack equivalent on $(G,S)$.
Based on the above-analyzed transformation result, that is, viewing $G||CE||BT(S)^{A}$ as a new plant and the sensor attacker $A$ as a new supervisor, we could naturally transform the attack model reduction problem to the supervisor reduction problem \cite{SW2004reduction}-\cite{SW2018reductionJ}, to generate the desired $A'$. Next, we shall define the following pieces of information:
\begin{itemize}
\setlength{\itemsep}{3pt}
\setlength{\parsep}{0pt}
\setlength{\parskip}{0pt}
    \item Let $En_{A}: Q_{a} \rightarrow 2^{\Sigma_{a}}$ with
        \[
        q \mapsto En_{A}(q) := \{\sigma \in \Sigma_{a}|\xi_{a}(q,\sigma)!\}
        \]
        be the ($A$-)enabled event set at state $q \in Q_{a}$.
    \item Let $D_{A}: Q_{a} \rightarrow 2^{\Sigma_{a}}$ with
        \[
        \begin{aligned}
        q \mapsto D_{A}(q) := \{\sigma \in \Sigma_{a}|\neg \xi_{a}(q,\sigma)! \wedge (\exists s \in L(G||CE||\\BT(S)^{A}))s\sigma \in L(G||CE||BT(S)^{A}) \wedge \xi_{a}(q_{a}^{init}, s) = q 
        \}
        \end{aligned}
        \]
        be the ($A$-)disabled event set at state $q \in Q_{a}$.
\end{itemize}
To obtain the ($A$-)enabled event set at state $q \in Q_{a}$, we just need to check the transition structure of $A$. To determine $D_{A}(q)$ for each state $q \in Q_{a}$, we can first compute the product $G||CE||BT(S)^{A}||A$, and then check in $G||CE||BT(S)^{A}||A$ each state tuple $(q_{G},q_{ce},q_{bs,a},q)$ associated with the state $q \in Q_{a}$. 

Let $\mathcal{R} \subseteq Q_{a} \times Q_{a}$ be a binary relation, where $(q, q') \in \mathcal{R}$ iff the following property hold:
\[
En_{A}(q) \cap D_{A}(q') = En_{A}(q') \cap D_{A}(q) = \varnothing
\]
We call $\mathcal{R}$ the \emph{binary compatibility relation} over $Q_{a}$ \cite{SW2018reduction}. This condition requires that no event enabled at one state can be disabled at the other state. 
For any two states satisfying $\mathcal{R}$, they may potentially be merged together, if their suffix behaviors are ``compatible'', which could be captured in the following definition. Let $I$ be a finite index set. 

\emph{Definition III.3} A collection $\mathcal{C} = \{(Q_{a,i},i)|Q_{a,i} \subseteq Q_{a} \wedge i \in I\}$ is a \emph{control congruence}  \cite{SW2004reduction} on $A$ if 
\begin{enumerate}[1)]
\setlength{\itemsep}{3pt}
\setlength{\parsep}{0pt}
\setlength{\parskip}{0pt}
    \item $\bigcup\limits_{i \in I}Q_{a,i} = Q_{a}$, $(\forall (Q_{a,i}, i), (Q_{a,j}, j) \in 2^{Q_{a}} \times I) i = j \Rightarrow Q_{a,i} = Q_{a,j} \wedge i \neq j \Rightarrow Q_{a,i} \cap Q_{a,j} = \varnothing$
    \item $(\forall i \in I) Q_{a,i} \neq \varnothing \wedge (\forall q, q' \in Q_{a,i}) (q, q') \in \mathcal{R}$
    \item $(\forall i \in I) (\forall \sigma \in \Sigma_{a}) (\exists j \in I) [(\forall q \in Q_{a,i}) \xi_{a}(q, \sigma)! \Rightarrow \xi_{a}(q, \sigma) \in Q_{a,j}]$
\end{enumerate}

Given a control congruence $\mathcal{C} = \{(Q_{a,i},i)|Q_{a,i} \subseteq Q_{a} \wedge i \in I\}$ on $A$, we shall construct an induced sensor attacker $A_{\mathcal{C}} = (I, \Sigma_{a}, \kappa, i_{0})$, where
\begin{itemize}
\setlength{\itemsep}{3pt}
\setlength{\parsep}{0pt}
\setlength{\parskip}{0pt}
    \item $i_{0} \in I$ such that $q_{a}^{init} \in Q_{a,i_{0}}$
    \item $\kappa: I \times \Sigma_{a} \rightarrow I$ is the (partial) transition function such that for each $i \in I$ and $\sigma \in \Sigma_{a}$, $\kappa(i, \sigma) := j$ if $j$ is chosen to satisfy the following property: there exists $q \in Q_{a,i}$ such that $\xi_{a}(q, \sigma) \in Q_{a,j}$ and 
    \[
    (\forall q' \in Q_{a,i})\xi_{a}(q',\sigma)! \Rightarrow \xi_{a}(q',\sigma) \in Q_{a,j}
    \]
    otherwise, $\kappa$ is not defined.
\end{itemize}

\emph{Theorem III.1:} Given a sensor attacker $A = (Q_{a}, \Sigma_{a} = \Sigma \cup \Sigma_{s,a}^{\#}, \xi_{a}, q_{a}^{init})$ for a closed-loop system, consisting of a plant $G = (Q, \Sigma, \xi, q_{0}, Q_{d})$ and a supervisor $S = (Q_{s}, \Sigma, \xi_{s}, q_{s}^{init})$, let $\mathcal{C} = \{(Q_{a,i},i)|Q_{a,i} \subseteq Q_{a} \wedge i \in I\}$ be a control congruence on $A$, and $A_{\mathcal{C}}$ be an induced sensor attacker from $\mathcal{C}$. Then $A_{\mathcal{C}}$ is attack equivalent to $A$.

\emph{Proof:}  
To prove $A_{\mathcal{C}}$ is attack equivalent to $A$, we need to show $L(G||CE||BT(S)^{A}||A) = L(G||CE||BT(S)^{A}||A_{\mathcal{C}})$ and $L_{m}(G||CE||BT(S)^{A}||A) = L_{m}(G||CE||BT(S)^{A}||A_{\mathcal{C}})$. Based on the above-analyzed transformation result, that is, viewing $G||CE||BT(S)^{A}$ as a new plant and the sensor attacker $A$ as a new supervisor, we just need to prove the following result:  Given a supervisor $A = (Q_{a}, \Sigma_{a} = \Sigma \cup \Sigma_{s,a}^{\#}, \xi_{a}, q_{a}^{init})$ for a plant $G||CE||BT(S)^{A}$, and a control congruence $\mathcal{C} = \{(Q_{a,i},i)|Q_{a,i} \subseteq Q_{a} \wedge i \in I\}$ on $A$, an induced supervisor $A_{\mathcal{C}}$ from $\mathcal{C}$ is control equivalent to $A$. To prove it, we could directly follow the proof of \emph{Proposition 2.1} in \cite{SW2004reduction}. \hfill $\blacksquare$

To accomplish the sensor attack model reduction, let 
\[
\mathcal{C} := \{[q] \subseteq Q_{a}| q \in Q_{a} \wedge q \in [q]\}
\]
be a control congruence on $A$, initially set to be
\[
(\forall q \in Q_{a}) [q] := \{q\}
\]
then we could adopt a polynomial-time algorithm, named as \emph{Reduction algorithm (RA)}, in \cite{SW2004reduction} to generate a new control congruence, whose induced sensor attacker $A_{\mathcal{C}}$ is unique \cite{SW2004reduction}. 


\section{EXAMPLE}
\label{sec:example}

\begin{figure}[htbp]
\centering
\subfigure[]{
\begin{minipage}[t]{0.43\linewidth}
\centering
\includegraphics[height=0.75in]{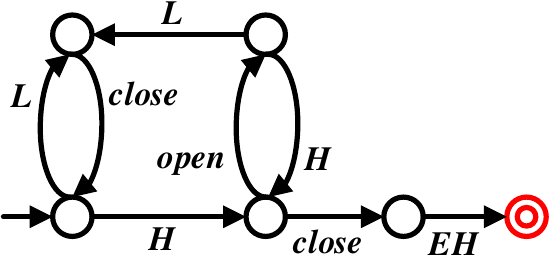}
\end{minipage}
}
\subfigure[]{
\begin{minipage}[t]{0.5\linewidth}
\centering
\includegraphics[height=1.15in]{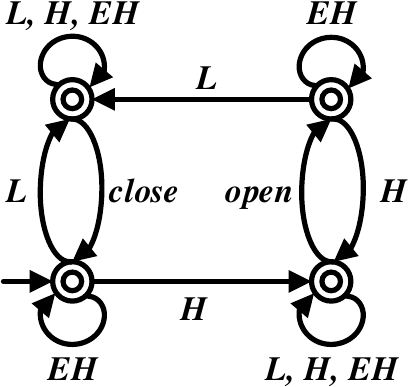}
\end{minipage}
}

\centering
\caption{(a) Plant $G$. (b) Supervisor $S$.}
\label{fig:G and S}
\end{figure}

\begin{figure}[htbp]
\begin{center}
\includegraphics[height=4.6cm]{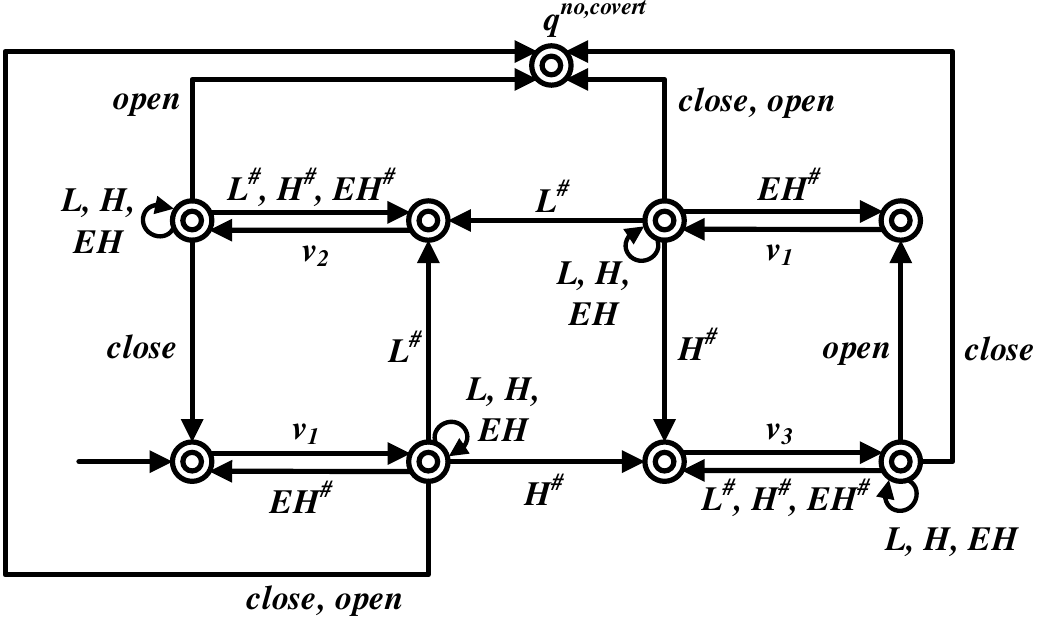}   
\caption{Bipartite supervisor under attack $BT(S)^{A}$}
\label{fig:bipartite supervisor under attack}
\end{center}        
\end{figure}

\begin{figure}[htbp]
\centering
\subfigure[]{
\begin{minipage}[t]{0.43\linewidth}
\centering
\includegraphics[height=1.4in]{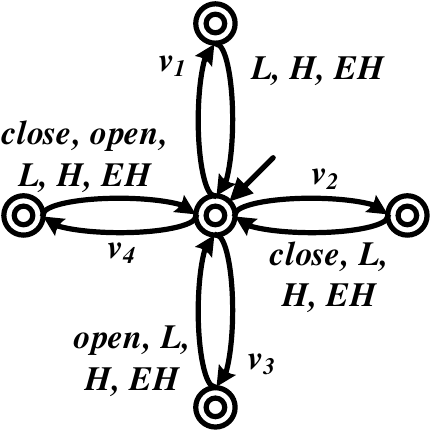}
\end{minipage}
}
\subfigure[]{
\begin{minipage}[t]{0.5\linewidth}
\centering
\includegraphics[height=0.4in]{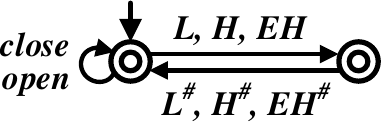}
\end{minipage}
}

\centering
\caption{(a) Command execution automaton $CE$. (b) Sensor attack constraints $AC$.}
\label{fig:CE and AC}
\end{figure}

We adopt the water tank example of \cite{LS20J} in this work. The system has a constant supply rate, a tank, and a control valve at the bottom of the tank controlling
the outgoing flow rate. We assume that the valve can only be fully open or fully closed, corresponding to two events: $open$ and $close$. The water level can be measured, whose value can trigger some predefined events that represent the water levels: low ($L$), high ($H$), and extremely high ($EH$). The models of the plant $G$ and the supervisor $S$ are shown in Fig. \ref{fig:G and S}. (a) and (b), respectively. The state marked red is the damage state of $G$. Double-edged circles are marker states. $\Sigma_{o} = \Sigma = \{L,H,EH,close,open\}$. $\Sigma_{c} = \{close,open\}$. $\Sigma_{o,a} = \{L,H,EH,close,open\}$, $\Sigma_{s,a} = \{L,H,EH\}$. $\Gamma = \{v_1,v_2,v_3,v_4\}$, where $v_1 = \{L,H,EH\}$, $v_2 = \{L,H,EH,close\}$, $v_3 = \{L,H,EH,open\}$ and $v_4 = \{L,H,EH,close,open\}$. The bipartite supervisor under attack $BT(S)^{A}$, command execution automaton $CE$ and sensor attack constraints $AC$ are shown in Fig. \ref{fig:bipartite supervisor under attack}, Fig. \ref{fig:CE and AC}. (a) and Fig. \ref{fig:CE and AC}. (b), respectively.

\begin{figure}[htbp]
\begin{center}
\includegraphics[height=2.9cm]{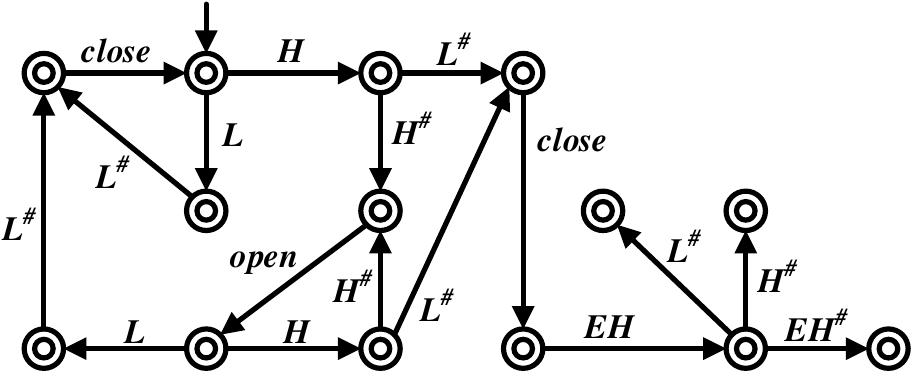}   
\caption{Sensor attacker $A$ (14 states)}
\label{fig:sensor attacker}
\end{center}        
\end{figure}

\begin{figure}[htbp]
\begin{center}
\includegraphics[height=1.7cm]{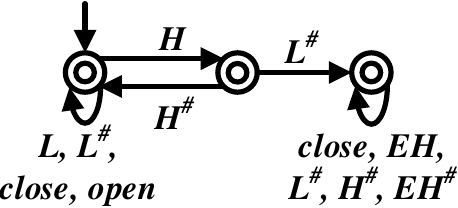}   
\caption{Reduced sensor attacker $A'$ (3 states)}
\label{fig:Reduced sensor attacker}
\end{center}        
\end{figure}

To show the effectiveness of the developed attack model reduction technique, we take the covert damage-nonblocking sensor attacker $A$ synthesized in \cite{LS20J}, shown in Fig. \ref{fig:sensor attacker}, as an instance. The reduced attack model $A'$ is shown in Fig. \ref{fig:Reduced sensor attacker}. By comparison, on one hand, $A$ has 14 states while the reduced model $A'$ only has 3 states, thus, the compression ratio \cite{SW2004reduction} is $\frac{14}{3} \approx 4.67$, which numerically verifies the effectiveness of our attack model reduction method. On the other hand, it can be seen that the attack model $A$ is complex and not easy for designers to grasp the attack logic. However, after the attack model reduction, the reduced model $A'$ clearly and intuitively reveals the key observation sequence that could induce the damage infliction, that is, to lead to the occurrence of the event $EH$ (water level becomes extremely high), once the sensor attacker receives $H$, it should alter it into $L^{\#}$. Then, based on the model $S$, the supervisor would issue the control command $v_{2} = \{L,H,EH,close\}$ under the fake sensor information $L^{\#}$, and the valve is closed, resulting in that the water level finally becomes extremely high, meaning the damage infliction goal is achieved. 

\section{CONCLUSIONS}
\label{sec:conclusion}

This work investigates how to identify the system vulnerability under sensor attack via attack model reduction technique. By constructing appropriate component models, we have shown that the attack model reduction problem could naturally be reduced to the well-studied supervisor reduction problem, which allows many existing tools to be used without devoting efforts to develop new techniques. This reduced attack model could provide a simplified attack logic, which discloses the key observation sequences resulting in the damage infliction, and thus guide system designers to fix bugs and improve security level. Furthermore, following the same spirit of this work, a distributed attack strategy that ensures attack equivalence can be developed, similar to the strategy of supervisor localization \cite{wonham2015supervisory,CW2010}, which might simplify the attack and make it more covert.











\end{document}